\documentclass{natureprintstyle}

\usepackage{stfloats}
\usepackage{float}
\usepackage{amsmath}
\usepackage{amssymb}
\usepackage{marvosym} 
\usepackage{multicol} 
\usepackage{siunitx}

\usepackage{tabularx} 
\newcolumntype{Y}{>{\centering\arraybackslash}X} 

\usepackage{diagbox} 
\usepackage{multirow} 

\usepackage{graphicx}
\graphicspath{{.}{Figures/}}

\usepackage[modulo]{lineno}
\modulolinenumbers[1]

\usepackage{color}
\definecolor{extended}{cmyk}{0.50,0.90,0,0.35}

\usepackage{fancyhdr}
\newcommand{\sigRatio} {{\sigma_\text{D}} / { \left(2 \sigma_\text{H}\right) }}

\newcommand{\minitab}[2][c]{\begin{tabular}{@{\extracolsep{-1pt}}c@{\extracolsep{-1pt}}}{#1}#2\end{tabular}} 

\usepackage{authoraftertitle}
\def\MyMaketitle{%
  \newpage\spacing{0.5}\setlength{\parskip}{3pt}%
    {\scalefont{3.0}\noindent\sloppy%
        \begin{flushleft}\bfseries\MyTitle\end{flushleft} \par}%
    {\scalefont{1.0}\noindent\sloppy \MyAuthor \par}%
}
\renewenvironment{affiliations}{%
    \let\olditem=\item
    \rule{\textwidth}{0.4pt}\\*[6pt]
    \raggedright
    \renewcommand\item[1][]{$^{\arabic{enumi}}$\stepcounter{enumi}}
    \setcounter{enumi}{1}%
    \setlength{\parindent}{0in}%
    \sffamily\sloppy%
    \scalefont{0.75}
    }{\let\item=\olditem}

\newfloat{affiliationsFloat}{b}{}[section] 

\newenvironment{myAuthors}{%
    \rule{\textwidth}{0.4pt}\\*[6pt]
    \let\olditem=\author
    \raggedright
    \renewcommand\author{}
    \setlength{\parindent}{0in}%
    \sffamily\sloppy%
    \scalefont{1.0}
    }{\let\author=\olditem}
    
\renewenvironment{abstract}{%
    \rule{\textwidth}{0.4pt}\\*[6pt]
    \setlength{\parindent}{0in}%
    \setlength{\parskip}{0in}%
    \scalefont{1.25}
        }{\par}

\makeatletter
\let\saved@includegraphics\includegraphics
\AtBeginDocument{\let\includegraphics\saved@includegraphics}

\makeatother

\newcommand\pstrut{\rule[-1ex]{0pt}{3.5ex}}

\title{The Asymmetry of Antimatter in the Proton}

\begin{document}

\widowpenalty=0
\clubpenalty=0
\onecolumn

\MyMaketitle

\begin{flushright}
\begin{minipage}{0.70\textwidth}
\raggedright
{\bf 
\begin{myAuthors}
\author{J.~Dove$^1$,}
\author{B.~Kerns$^1$,}
\author{R.~E.~McClellan$^{1,18}$}
\author{S.~Miyasaka$^2$,}
\author{D.~H.~Morton$^3$,} 
\author{K.~Nagai$^{2,4}$,}
\author{S.~Prasad$^1$,}
\author{F.~Sanftl$^2$,}
\author{M.~B.~C.~Scott$^3$,}
\author{A.~S.~Tadepalli$^{5,18}$,}
\author{C.~A.~Aidala$^{3,6}$,}
\author{J.~ Arrington$^{7,19},$}
\author{C.~Ayuso$^{3,20}$,}
\author{C.~L.~Barker$^8$,}
\author{C.~N.~Brown$^9$,}
\author{W.C.~Chang$^4$,}
\author{A.~Chen$^{1,3,4}$,}
\author{D.~C.~Christian$^{10}$,}  
\author{B.~P.~Dannowitz$^1$,}
\author{M.~Daugherity$^8$,}
\author{M.~Diefenthaler$^{1,18}$,}
\author{L.~El~Fassi$^{5,11}$,} 
\author{D.~F.~Geesaman$^{7,21}$,} 
\author{R.~Gilman$^5$,}
\author{Y.~Goto$^{12}$,}
\author{L.~Guo$^{6,22}$,}
\author{R.~Guo$^{13}$,}
\author{T.~J.~Hague$^8$,}
\author{R.~J.~Holt$^{7,23}$,} 
\author{D.~Isenhower$^8$,} 
\author{E.~R.~Kinney$^14$,}
\author{N.~Kitts$^8$,}
\author{A.~Klein$^6$,}
\author{D.~W.~Kleinjan$^6$,}
\author{Y.~Kudo$^{15}$,}
\author{C.~Leung$^1$,}
\author{P.-J.~Lin$^{14}$,}
\author{K.~Liu$^6$,} 
\author{M.~X.~Liu$^6$,} 
\author{W.~Lorenzon$^3$,}
\author{N.~C.~R.~Makins$^1$,}
\author{ M.~Mesquita de Medeiros$^7$,}
\author{P.~L.~McGaughey$^6$,}
\author{Y.~Miyachi$^{15}$,} 
\author{I.~Mooney$^{3,24}$,}
\author{K.~Nakahara$^{16,25}$,}
\author{K.~Nakano$^{2,12}$,}
\author{S.~Nara$^{15}$,} 
\author{J.-C.~Peng$^1$,}
\author{A.~J.~Puckett$^{6,26}$,}
\author{B.~J.~Ramson$^{3,27}$,} 
\author{P.~E.~Reimer$^{7}$\Letter,}
\author{J.~G.~Rubin$^{3,7}$,}
\author{S.~Sawada$^{17}$,} 
\author{T.~Sawada$^{3,28}$,}
\author{T.-A.~Shibata$^{2,29}$,} 
\author{D.~Su$^4$,}
\author{M.~Teo$^{1,30}$,}
\author{B.~G Tice$^7$,}
\author{R.~S.~Towell$^8$,}
\author{S.~Uemura$^{6,31}$,}
\author{S.~Watson$^8$,}
\author{S.~G.~Wang$^{4,13,32}$,}
\author{A.~B.~Wickes$^6$,}
\author{J.~Wu$^{10}$,}
\author{Z.~Xi$^8$,}
\author{Z.~Ye$^7$} 
\end{myAuthors}
}

\vspace{12pt}

\begin{abstract}The fundamental building blocks of the proton, quarks and gluons, have been known for decades. However, we still have an incomplete theoretical and experimental understanding of how these particles and their dynamics give rise to the quantum bound state of the proton and its physical properties, such as for example its spin.\cite{Ji:2020ena} The two up and the single down quarks that comprise the proton in the simplest picture account only for a few percent of the proton mass, the bulk of which is in the form of quark kinetic and potential energy and gluon energy from the strong force.\cite{PhysRevLett.121.212001} An essential feature of this force, as described by quantum chromodynamics, is its ability to create matter-antimatter quark pairs inside the proton that exist only for a very short time. Their fleeting existence makes the antimatter quarks within protons difficult to study, but their existence is discernible in reactions where a matter-antimatter quark pair annihilates. In this picture of quark-antiquark creation by the strong force, the probability distributions as a function of momentum for the presence of up and down antimatter quarks should be nearly identical, since their masses are quite similar and small compared to the mass of the proton.\cite{ROSS1979497} In the present manuscript, we show evidence from muon pair production measurements that these distributions are significantly different, with more abundant down antimatter quarks than up antimatter quarks over a wide range of momentum. These results revive interest in several proposed mechanisms as the origin of this antimatter asymmetry in the proton that had been disfavored by the previous results\cite{Towell:2001nh} and point to the future measurements that can distinguish between these mechanisms.
\end{abstract}

\end{minipage}
\end{flushright}

\begin{affiliationsFloat}
\begin{affiliations}
\item Department of Physics, University of Illinois at Urbana-Champaign, Urbana, IL, USA.
\item Department of Physics, School of Science, Tokyo Institute of Technology, Meguro-ku,Tokyo, Japan.
\item Randal Laboratory of Physics, University of Michigan, Ann Arbor, Michigan, USA.
\item Institute of Physics, Academia Sinica, Taipei, Taiwan.
\item Department of Physics and Astronomy, Rutgers, The State University of New Jersey, Piscataway, New Jersey, USA.
\item Physics Division, Los Alamos National Laboratory, Los Alamos, New Mexico, USA.
\item Physics Division, Argonne National Laboratory, Lemont, Illinois, USA.
\item Department of Engineering and Physics, Abilene Christian University, Abilene, Texas, USA.
\item Accelerator Division, Fermi National Accelerator Laboratory, Batavia, Illinois, USA.
\item Particle Physics Division Fermi National Accelerator Laboratory, Batavia, Illinois, USA.
\item Department of Physics and Astronomy, Mississippi State University, Mississippi State, MS, USA. 
\item RIKEN Nishina Center for Accelerator-Based Science, Wako, Saitama, Japan.
\item Department of PHysics,National Kaohsiung Normal University, Kaohsiung City, Taiwan.
\item Department of Physics, University of Colorado, Boulder, Colorado, USA. 
\item Department of Physics, Yamagata University, Yamagata City, Yamagata, Japan.
\item Department of Physics, University of Maryland, College Park, Maryland, USA.
\item Institute of Particle and Nuclear Studies, KEK, High Energy Accelerator Research Organization, Tsukuba, Ibaraki, Japan.
\item Present address: Experimental Nuclear Physics Division, Thomas Jefferson National Accelerator Facility, Newports News, IL, USA.
\item Present address: Physics Division, Lawrence Berkeley National Laboratory, Berkeley, California, USA.
\item Present address: Department of Physics and Astronomy, Mississippi State University, Mississippi State, MS, USA.
\item Present address: Argonne Associate of Global Empire LLC, Lemont IL, USA.
\item Present address: Department of Physics, Florida International University, Miami, Florida, USA.
\item Present address: Kellogg Radiation Laboratory, California Institute of Technology, Pasadena, CA, USA.
\item Present address: Department of Physics, Wayne State University, Detroit, MI, USA.
\item Present address: Stanford Linear Accelerator Center, Menlo Park, CA, USA.
\item Present address: Department of Physics, University of Connecticut, Storrs, CT, USA.
\item Present address: Neutrino Division, Fermi National Accelerator Laboratory,
\item Present address: Department of Physics, Osaka City University, Sumiyoshi-ku, Osaka City, Osaka 558-8585, Japan.
\item Present address: Department of Physics, College of Science and Technology, Nihon University, Tokyo, Japan.
\item Present address: Department of Physics, Stanford University, Stanford, CA, USA.
\item Present address: Department of Physics, Tel Aviv University, Tel Aviv, Israel.
\item Present address: Accelerator Systems Division, Argonne National Laboratory, Lemont, IL, USA.

\Letter email:reimer@anl.gov
\end{affiliations}
\end{affiliationsFloat}
\sloppy
 
\begin{multicols}{2}
The structure of the proton is a prototypical example of a strongly coupled and correlated system with the quarks and gluons interacting according to quantum chromodynamics (QCD). At large energy and momentum scales, the interaction is relatively weak, while at lower energy scales the picture is still clouded by the increasingly strong interaction. The original quark model, in which the proton consists of two up quarks (u) and one down (d) quark, has an appealing simplicity, but experiments that measure the distributions of quarks as a function of the fractional momentum of the proton that they carry ($x$) have revealed a rich structure with additional quarks, antimatter quarks (antiquarks), and gluons beyond the minimal three quark Fock state. These additional quarks and antiquarks are referred to as sea quarks. Collectively the quarks and gluons are referred to as partons. It is not possible to identify any individual up or down quark as a sea or valence quark, but antiquarks and strange quarks must belong to the sea and so their study promises to reveal new information about the structure of the proton.  Even prior to QCD, hadronic models emphasized the importance of the presence of mesons (e.g., Ref.~[\citen{Fermi:1947}]) and therefore, as was realized later, antiquarks, in the physical state of a proton or neutron. Despite this, the initial naive expectation was that the sea was formed predominantly by gluons splitting into quark-antiquark pairs. Indeed, several authors assumed that at some low momentum scale the sea quarks and gluons vanish and  all the sea and the glue at high momentum scales are generated by gluon radiation and then gluon splitting. They were able to describe successfully the existing data in the late 1970's (e.g., Ref. [\citen{Gluck:1977ah, Parisi:1976fz}]).

In the early 1990's the New Muon Collaboration (NMC) reported measurements of the deep inelastic structure functions ($F_2$) of hydrogen (H) and deuterium (D)\cite{Amaudruz:1991at,Arneodo:1994sh} at $0.004 < x<0.8$. The cross section for deep inelastic scattering measures the charge--squared-weighted sum of the quark and antiquark distributions, in this case at an average scale of 4 $\left(\text{GeV}/c\right)^2$. The integrals of the parton distributions of the proton (p) and neutron (n) were assumed to have charge symmetry, $\int_0^1 u_p(x) dx=\int_0^1 d_n(x) dx$, with similar integrals for the other quark flavors, and nuclear effects in deuterium were assumed to be small $\left(F_2^D = F_2^p + F_2^n\right)$.  In that case their measurements and their estimate of the unmeasured region led NMC to conclude
\begin{eqnarray}
 \lefteqn{\int_0^1 \frac{dx}{x}  \left[F_{2}^p(x) - F_{2}^n(x)\right]} \\ \nonumber
 & = &  \frac{1}{3} + \frac{2}{3} \int_0^1dx \left[\bar{u}(x) - \bar{d}(x) \right] \\ \nonumber
 & = &  0.235 \pm 0.026, 
\end{eqnarray}
and thus the integral of $\bar{d}(x)$ is greater than the integral of $\bar{u}(x)$:
\begin{equation} 
 \int_0^1dx \left[\bar{d}(x) - \bar{u}(x)\right]= 0.147 \pm 0.039,
\end{equation} 
where $\bar{u}(x)$ and $\bar{d}(x)$ are the distributions of up and down antiquarks in the proton respectively. 

The Drell-Yan process in hadron-hadron collisions is a reaction where a quark and an antiquark annihilate into a virtual photon and that virtual photon decays into a lepton-antilepton pair.\cite{Drell:1970wh} One can isolate the antiquark distributions from the Drell-Yan cross section making use of this property.  At lowest order the Drell-Yan cross section is given by
\begin{equation}
\frac{d^2\sigma}{dx_b\,dx_t} = \frac{4\,\pi\,\alpha^2}{9\,s\,x_b\,x_t}\,\sum_q\,e_q^2
\left[q(x_b)\bar{q}(x_t) + \bar{q}(x_b)q(x_t)\right],
\end{equation}
where $x_b$ and $x_t$ are the momentum fractions of the beam and target partons participating in the reaction, $e_q$ is the electrical charge of quark flavor $q$, $\alpha$ is the fine structure constant, and $s$ is the square of the center of mass energy of the beam and target. In a Drell-Yan measurement at CERN, the  NA51 collaboration confirmed\cite{Baldit:1994jk} that $\bar{d}(x)$ is larger than $\bar{u}(x) $ at an average $x$ value of 0.18.  

When a Drell-Yan experiment is performed with a proton beam and kinematic acceptance that selects events with $x_b$ in the valence-quark dominated region and with Feynman $x$, $x_F \equiv x_b - x_t \gg 0$, the first term in equation 3 dominates. The charge-squared weighting and the fact that $u_v(x)$ is approximately $2 d_v(x)$ for the valence quark distributions of the proton beam means that the measurement is, by a factor of approximately eight, more sensitive to $\bar{u}$ quarks in the target than $\bar{d}$. The renormalization and factorization scales for the extraction of parton distributions are usually chosen as the mass squared of the virtual photon times the speed of light squared,  $M^2 c^2= x_b\,x_t\,s/c^2 - P_T^2$ where $P_T^2$ is the square of the transverse momentum of the virtual photon and is usually small compared to $M^2 c^2$. Using charge symmetry,\cite{Lon:1998} to relate proton and neutron parton distributions, 
$\left(\vphantom{\bar d_n} u_p(x) = d_n(x) \right.$, 
$d_p(x) =  u_n(x) $, 
$ \bar{u}_p(x)  = \bar{d}_n(x) $,
$\left.\bar{d}_p(x) =  \bar{u}_n(x) \right)$,
as is assumed by almost all the global parton distribution fits, and assuming that the nuclear corrections in the deuteron are small, as supported by calculations,\cite{KamLee:2012,Ehlers:2014} the ratio of the Drell-Yan cross section on a deuterium target to that on a hydrogen target, 
$\sigma_D / \sigma_H  \approx \left(\sigma_p + \sigma_n\right) / \sigma_p \approx 1 + \bar{d}_p(x_t) / \bar{u}_p(x_t)$, almost directly measures $\bar{d}(x_t)/\bar{u}(x_t)$.

\begin{figure*}[ht]
	\centering
 		\caption{ \linespread{1}\selectfont{} {\bf Ratios of $\mathbf{\sigma_D}$ to $\mathbf{2\sigma_H}$.}\label{fig:crRatio} }
		\includegraphics[width=0.75\textwidth]{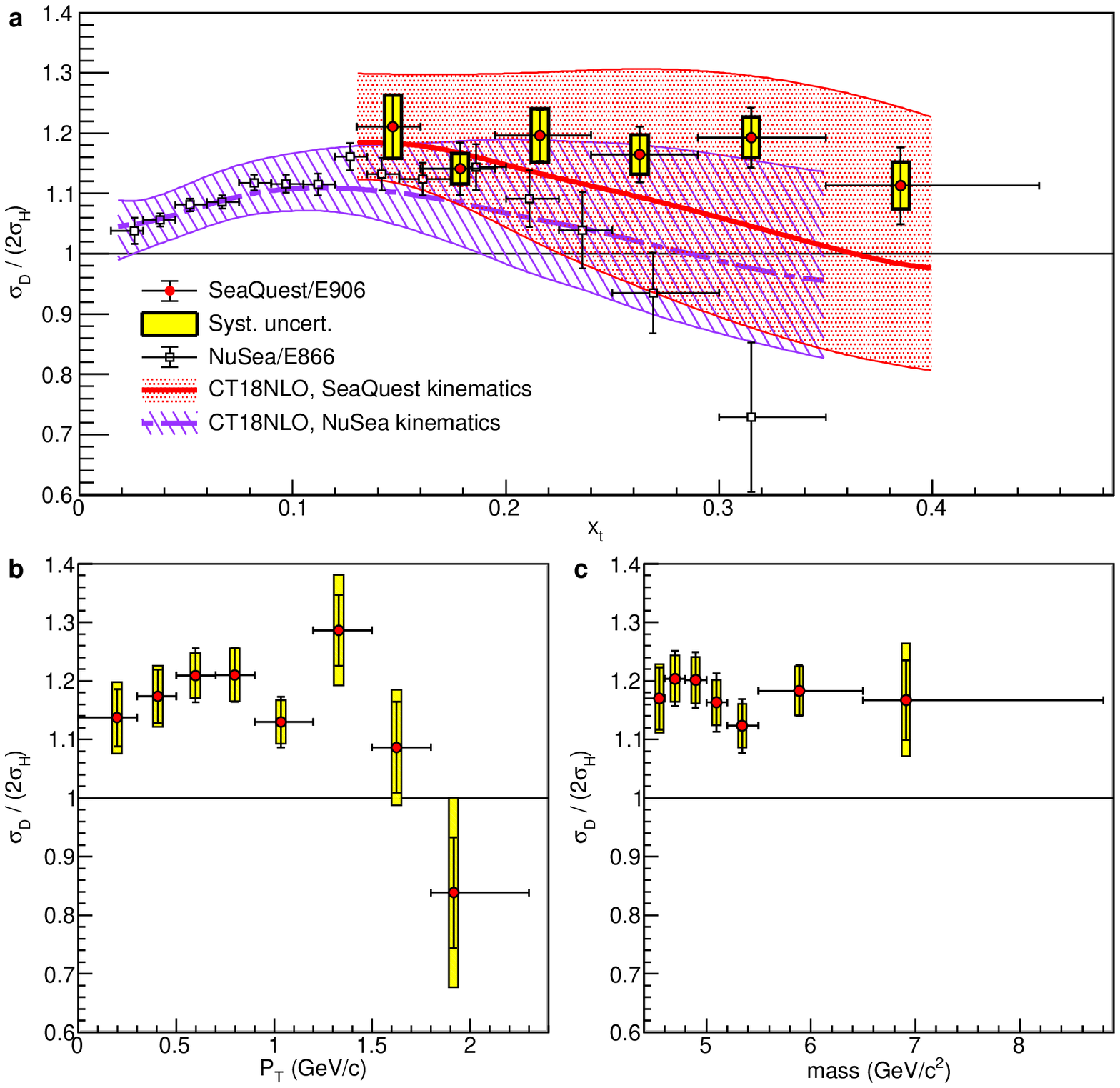}

	\flushleft
		\small{Ratios of $\sigma_D(x_t)$ to $2\sigma_H(x_t)$ (Upper plot, red filled circles) with their statistical (vertical bars) and systematic (yellow boxes) uncertainties as a function of $x_t$, and (lower plots) as functions of transverse momentum, $P_T$, and mass, $M$, of the virtual photon. The cross section ratios are defined as the ratio of luminosity-corrected yields from the hydrogen and deuterium targets. Also shown (open black squares) are the results of the NuSea experiment\cite{Towell:2001nh} for the $x_t$ dependence with statistical uncertainties only.  NuSea also reports an overall 1\% common systematic uncertainty. The mass scale of the NuSea data is up to 50\% larger than that of the SeaQuest data and the distributions in the other kinematic variable, $x_b$, are slightly different due to the differing beam energies and acceptances of the experiments. These differences imply that the cross section ratios do not have to be identical.  This is demonstrated by the red solid and violet long-short dashed curves representing NLO calculations of the cross section ratio at SeaQuest and NuSea kinematics using CT18 parton distributions~\cite{Hou:2019efy}.  The horizontal bars on the data points indicate the width of the bins.}
\end{figure*}

\begin{figure*}
	\centering
		\caption{\linespread{1}\selectfont{}{\bf Ratios of $\mathbf{\bar{d}(x)}$ to $\mathbf{\bar{u}(x)}$.} \label{fig:dbub} }
		\includegraphics[width=0.75\textwidth]{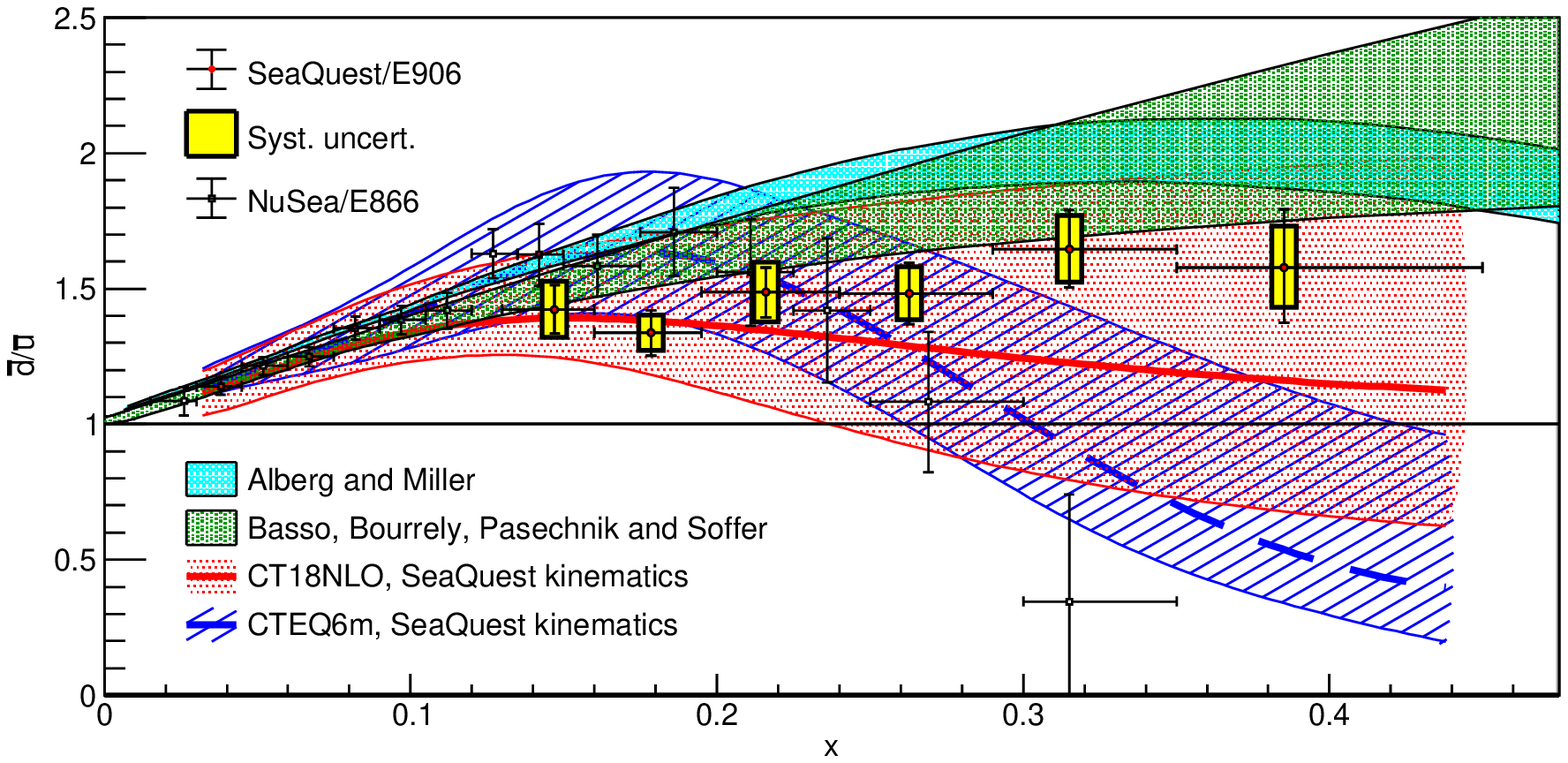}

	\flushleft
		\small{Ratios of $\bar{d}(x)$ to $\bar{u}(x)$ in the proton (red filled circles) with their statistical (vertical bars) and systematic (yellow boxes) uncertainties extracted from the present data based on next-to-leading order calculations of the Drell-Yan cross sections. Also shown in the open black squares are the results obtained by the NuSea experiment with statistical and systematic uncertainties added in quadrature\cite{Towell:2001nh}. The cyan band shows the predictions of the meson-baryon model of Alberg and Miller\cite{Alberg:2017ijg} and the green band shows the predictions of the statistical parton distributions of Basso, Bourrely, Pasechnik and Soffer\cite{Bourrely:2016}. The red solid (blue dashed) curves show the calculated ratios of $\bar{d}(x)$ to $\bar{u}(x)$ with  CT18\cite{Hou:2019efy}  (CTEQ6\cite{Pumplin:2002vw}) parton distributions at the scales of the SeaQuest results.  The horizontal bars on the data points indicate the width of the bins.}
\end{figure*}

The Fermilab NuSea/E866 collaboration\cite{Towell:2001nh} (whose results are displayed in Fig.~\ref{fig:crRatio} and Fig.~\ref{fig:dbub}) was able to measure the $x_t$ dependence of the $\sigRatio$ ratio with an 800 GeV proton beam in the kinematic range $0.015<x_t<0.35$ and by extrapolating the results to $x_t=0$ and  $x_t=1$ obtained a value of the $\int_{0}^{1} dx \big[\bar{d}(x) - \bar{u}(x)\big] = 0.118 \pm 0.012$ at an average scale of $54 \left(\text{GeV}/c\right)^2$. The HERMES collaboration also measured part of this integral and obtained results\cite{Ackerstaff:1998} consistent with NMC and NuSea. One feature of the NuSea results, with admittedly limited statistics, is the suggestion that the ratio of $\bar{d}\left(x\right) / \bar{u}\left(x\right)$ began to decrease  for $x>0.2$, reaching a value of $\bar{d}(x) / \bar{u}(x)$ = $0.35 \pm 0.40$ at $x=0.31$, as seen in Fig.~\ref{fig:dbub}.  

There are a variety of mechanisms that may account for the antiquark flavor asymmetry of the proton; recent reviews include Ref. [\citen{Chang:2014jba , Geesaman:2019}].  Pauli blocking\cite{Field:1976ve} may lead to a flavor asymmetry as the extra $u$ valence quark Pauli blocks some $u$-$\bar u$ pairs from forming, but the $x$ dependence and even the sign of this mechanism are debated in the literature~\cite{Ross:1978xk ,Steffens:1996bc}. A related approach involves statistical models.\cite{Bourrely:2015kla,Bourrely:2016} Another class including chiral soliton models\cite{Pobylitsa:1999} and meson-baryon models emphasize mesonic degrees of freedom in proton structure.\cite{Thomas:1983fh,Alberg:2012wr,Alberg:2017ijg}  These latter models (statistical, chiral soliton, and meson-baryon) each attempt to describe the entire non-perturbative composition of the proton.  A common feature of these models is a rise in the $\bar d/\bar u$ flavor asymmetry with $x$. While at low $x$ this behavior reproduces the NuSea data, none of these models are able to reproduce the fall-off at higher $x$ observed by NuSea.  The only {\em ab initio} technique for calculating the parton distributions of the proton is lattice QCD (recently reviewed by Lin {\it et al.}\cite{Lin:2017snn}). At this time, the lattice results for both quarks and antiquarks are still not in quantitative agreement with global fits of parton distributions to experimental data and the systematic errors are still being evaluated. 

The SeaQuest experiment (E906) at Fermi National Accelerator Laboratory (Fermilab) was designed to investigate the flavor asymmetry at higher $x_t$ values than NuSea with the newly constructed experimental apparatus that is described in detail in Ref.~[\citen{SeaquestNIM:2019}]. With a proton beam at an energy of \SI{120}{\GeV}, liquid hydrogen and deuterium targets, and a 10 T-m focusing magnet after the target region, the experiment was optimized for the study of the target antiquarks in the intermediate region, with $x_t$ around $0.3$, by detecting the $\mu^+\mu^-$ pairs from decays of the virtual photons produced in the Drell-Yan process. The proton beam was extracted from the Fermilab Main Injector using slow-spill extraction for \SI{4}{\second} every \SI{60}{\second}. The microstructure of the beam consisted of  \SI{1}{\ns} long bunches of approximately 0 to 80,000 protons  at a \SI{53}{\MHz} repetition rate. About $6 \times 10^{12}$ protons were incident on the target in the four second long extraction period. A Cherenkov detector, installed in the beam line, measured the number of protons for each  bunch and allowed high intensity bunches (usually greater than 64,000 protons) to be vetoed. 

\begin{table*}[t]
	\caption{ \linespread{1}\selectfont{}{\bf Ratios of $\mathbf{\sigma_D}$ to $\mathbf{2\sigma_H}$.} \label{tab:crRatio}} \vspace{6pt}
	\begin{tabularx}{\textwidth} { Y Y Y Y Y r @{ $\pm$ }c@{ $\pm$ }l  Y }  \hline

$x_t$ bin & $\langle x_t\rangle$ & $\langle x_b\rangle$ & $\langle M\rangle$ &$\langle P_T\rangle$ & $\sigRatio$ & stat. & syst.  & $\delta x_t$ \rule{0pt}{1.1em}\\ 
                 &             &               & (GeV/c$^2$) & (GeV/c)   & \multicolumn{3}{c} {} & \pstrut \\ \hline
$  0.130-0.160 $ & $ 0.147 $ & $ 0.688 $ & $ 4.71 $ & $ 0.651 $ & $ 1.211 $ & $ 0.052 $ & $ 0.053 $ & $ 0.013 $ \pstrut \\ \hline
$  0.160-0.195 $ & $ 0.179 $ & $ 0.611 $ & $ 4.88 $ & $ 0.717 $ & $ 1.141 $ & $ 0.043 $ & $ 0.025 $ & $ 0.016 $ \pstrut \\ \hline
$  0.195-0.240 $ & $ 0.216 $ & $ 0.554 $ & $ 5.11 $ & $ 0.757 $ & $ 1.196 $ & $ 0.042 $ & $ 0.044 $ & $ 0.019 $ \pstrut \\ \hline
$  0.240-0.290 $ & $ 0.263 $ & $ 0.519 $ & $ 5.46 $ & $ 0.786 $ & $ 1.165 $ & $ 0.046 $ & $ 0.032 $ & $ 0.022 $ \pstrut \\ \hline
$  0.290-0.350 $ & $ 0.315 $ & $ 0.498 $ & $ 5.87 $ & $ 0.785 $ & $ 1.193 $ & $ 0.050 $ & $ 0.034 $ & $ 0.026 $ \pstrut \\ \hline
$  0.350-0.450 $ & $ 0.385 $ & $ 0.477 $ & $ 6.36 $ & $ 0.776 $ & $ 1.113 $ & $ 0.064 $ & $ 0.039 $ &$ 0.030 $ \pstrut \\ \hline
	\end{tabularx}

	\vspace{6pt}
	\small{Cross section ratios $\sigRatio$  binned in $x_t$ with their statistical and systematic uncertainties and the average values for the kinematic variables of each $x_t$ bin. The cross section ratios are defined as the ratio of luminosity-corrected yields from the hydrogen and deuterium targets. The final column is the experimental resolution in $x_t$ as determined by Monte Carlo simulations.}
	\\
	\hrule
\end{table*}

The data analysis procedure is described in the ``Methods Summary" section. The primary challenge in the data analysis consisted of rate dependent effects that were exacerbated by large fluctuations in the bunch beam intensity. The average duty factor, defined as the square of the average  intensity  divided by the average of the intensity squared during the beam spill, $\langle I \rangle^2/ \langle I^2 \rangle$, of the beam ranged between 20\% and 40\%.  These intensity variations had two primary consequences: first, a variation in the track reconstruction efficiency, and second, a change in the rate of accidental coincidences.   Rather than trying to separate and model all the rate dependent effects, the ratios of the yields on deuterium, $Y_D$, and hydrogen, $Y_H$, were analyzed by fitting the ratios as a function of $x_t$ and bunch intensity, $I$, to a functional form. For the final analysis, the form $\frac{Y_D\left(x_{t},I\right)} {2 Y_H\left(x_{t},I\right)} = R_{x_{t}} + a I + b  I^2 $, was chosen based on the Akaike information criterion.\cite{Akaike:1974} Here, $x_t$ is the bin average of each $x_t$ bin, $R_{x_{t}}$ is the fitted cross section ratio at zero intensity, $ \sigRatio$, and $a$ and $b$ are constants fitted to the entire $x_t$ range. The results from this form were compared with other functional forms such as $\frac{Y_D\left(x_{t},I\right)} {2 Y_H\left(x_{t},I\right)}= R_{x_{t}} + \left(a_{0}+a_{1}  x_t \right)  I + \left(b_{0} +b_{1} x_t\right) I^2 $ and the differences between the results with these two forms were taken as part of the $x_t$ dependent systematic error.  The only other significant systematic error was an overall 2\% uncertainty related to the relative beam normalization. The cross section ratios, $\sigRatio$, defined as the ratio of luminosity-corrected yields from the hydrogen and deuterium targets, measured by SeaQuest as a function of $x_t$, $M$ and $P_T$ are show in Fig.~\ref{fig:crRatio} and tabulated in Tab.~\ref{tab:crRatio}, along with the average $x_b$, $M$, and $P_T$. (The cross section ratios values for the $P_T$ and $M$ plots in Fig.~\ref{fig:crRatio} are given in Extended Data Tabs.~\ref{exttab:pt}, and \ref{exttab:mass}.) The average values  of $x_t$, $x_b$, $M$, and $P_T$ in each bin are the same for deuterium and hydrogen within uncertainties and not corrected for acceptance.  The statistical uncertainty on each data point is the uncertainty returned from the fit on the zero intensity $\left(R\right)$ parameter, where the uncertainty of the individual ratios only included the counting uncertainties of Drell-Yan yields on the hydrogen and deuterium targets. Since the parameters $a$ and $b$ are fit over all $x_t$ bins, the statistical uncertainty is correlated by 40\% - 70\% among $x_t$ bins. The covariance matrix is given in the Eq.~\ref{eq:correlation} of the Methods section. The systematic uncertainty is fully correlated among all $x_t$ bins. The observation that both $M$ and $P_T$ (for all but the very highest $P_T$ bins) distributions for deuterium and hydrogen have the same shapes helps to confirm that the acceptances are very similar for each target.  Also shown in Fig.~\ref{fig:crRatio} are the results as a function of $x_t$ from NuSea\cite{Towell:2001nh} and the cross section ratio calculated in next-to-leading order (NLO) with CT18\cite{Hou:2019efy} parton distribution. The NuSea results are at a different scale,  $54 \left(\text{GeV}/c\right)^2$, than the SeaQuest results $22\text{-}40\left(\text{GeV}/c\right)^2$. The cross section ratios depend on both $x_b$ and $x_t$, and due to the differing beam energies and acceptances, the $x_b$ distributions are slightly different for SeaQuest and NuSea, the effects of which are shown in Fig.~\ref{fig:crRatio} and Extended Data Fig.~\ref{extfig:nloPDF}. 

To extract $\bar{d}(x)/\bar{u}(x)$, next to leading order calculations of $\sigRatio$  were carried out starting from several  NLO global fits to the parton distributions (CT10\cite{Lai:2010vv}, CT14\cite{Dulat:2015mca}, CT18\cite{Hou:2019efy}, MMHT2014\cite{MMHT:2014}). No nuclear correction for deuterium was applied but a systematic uncertainty of  (0.5+ 5$x_t$)\% is included based on the range of deviation from unity found in Ref. [\citen{KamLee:2012, Ehlers:2014}]. Holding all other parton distributions fixed including the sum $\bar{d}(x) + \bar{u}(x)$, the ratio $\bar{d}(x)/\bar{u}(x)$ for each $x_t$ bin was varied until the results converged on the measured cross section ratios. The correlations of the statistical uncertainties of $\sigRatio$ were propagated through the $\bar{d}/\bar{u}$ extraction. The dependence on the value for the ratio $\bar{d}(x)/\bar{u}(x)$ above the measured $x$ region was estimated by varying this value from 1.0 to 0.5 and 2.0. The spread of the results due to the choice of initial parton distributions was always less than half of the statistical error.  Each $x_t$ bin was subdivided into multiple $x_b$ sub-bins.  The cross sections for hydrogen and deuterium were calculated separately for each sub-bin at the $\langle x_t\rangle$, $\langle x_b\rangle$, and $\langle M\rangle$ of that sub-bin and an acceptance weighted sum was used to determine the final cross section.  These distributions are given in Extended Data Tab.~\ref{exttab:acc}. Calculations using only one average $x_t$ and $x_b$ for each $x_t$ bin were less reliable.  It was also found that a leading order extraction of $\bar{d}(x)/\bar{u}(x)$ using leading order parton distributions and cross section calculations gave very similar results for the ratios compared to the NLO results.

\begin{table*}
	\caption{\linespread{1}\selectfont{}{\bf Ratios of $\mathbf{\bar{d}(x)}$ to $\mathbf{\bar{u}(x)}$.} 
 \label{tab:dbub} } 
 	\vspace{6pt}
	\begin{tabularx}{\textwidth}{crclY} 
 \hline
$\langle x\rangle$ & $\bar{d}(x)/\bar{u}(x)$ &\multicolumn{1}{@{ $\pm$ }c}{stat.}  & \multicolumn{1}{@{ $\pm$ }r}{syst.}  \pstrut \\ \hline
\multirow[c]{2}{*}{$ 0.147 $ }& \multirow[c]{2}{*}{$ 1.423 $} & $+0.089$ & $+0.104$ &\rule[0.5ex]{0pt}{2ex}\\
                                             &                                                &  $-0.089 $ &  $-0.103$  \rule[-1ex]{0pt}{3ex} \\ \hline
\multirow[c]{2}{*}{$ 0.179 $ } & \multirow[c]{2}{*}{$ 1.338 $ }  & $+0.083 $ &  $+0.065 $ \rule[0.5ex]{0pt}{2ex}\\  
                                             &                                                &  $-0.085 $ & $-0.065 $  \rule[-1ex]{0pt}{3ex} \\ \hline
\multirow[c]{2}{*}{$ 0.216 $ } & \multirow[c]{2}{*}{$ 1.487 $ } &  $+0.092 $ & $+0.111 $ \rule[0.5ex]{0pt}{2ex} \\
                                           &                                                 &  $-0.092 $ & $-0.110 $  \rule[-1ex]{0pt}{3ex} \\ \hline
\multirow[c]{2}{*}{$ 0.263 $ } & \multirow[c]{2}{*}{$ 1.482 $ } &  $+0.114 $ &  $+0.098  $ \rule[0.5ex]{0pt}{2ex} \\
                                           &                                               &  $-0.113 $ &  $-0.097 $  \rule[-1ex]{0pt}{3ex} \\ \hline
\multirow[c]{2}{*}{$ 0.315 $ } & \multirow[c]{2}{*}{$ 1.645 $ } & $+0.144 $ &  $+0.125$ \rule[0.5ex]{0pt}{2ex} \\
                                             &                                               & $-0.121 $ & $-0.121$  \rule[-1ex]{0pt}{3ex} \\ \hline
\multirow[c]{2}{*}{$ 0.385 $ } & \multirow[c]{2}{*}{$ 1.578 $ } &  $+0.214$ &  $+0.153$ \rule[0.5ex]{0pt}{2ex} \\
                                              &                                              & $-0.203$ &  $-0.148$  \rule[-1ex]{0pt}{3ex} \\ \hline
	\end{tabularx}
	
	\vspace{6pt}
	\small{Ratios of $\bar{d}(x)$ to $\bar{u}(x)$ with their upper and lower statistical and systematic uncertainties.  The analysis was based on the present cross section ratio data, and next-to-leading order calculations of the Drell-Yan cross sections using CT18 parton distributions for all except the ratio of $\bar{d}(x)$ to $\bar{u}(x)$. The systematic uncertainty is fully correlated among all $x$ bins.  The systematic uncertainty does not include a contribution from the choice of the base (CT18) pdf, which is small if added in quadrature to the other systematic uncertainties.}
	\\
	\hrule
\end{table*}

The resulting ratios of $\bar{d}(x)$ to $\bar{u}(x)$ distributions starting with the CT18 distributions are given in Table~\ref{tab:dbub} at the scale of each $x_t$ bin and displayed in Fig.~\ref{fig:dbub}, and compared there with the NuSea results.  The trends between the two experiments at higher $x_t$ are quite different. No explanation has been found yet for the differing results, even though there is a small overlap in the members of the NuSea and SeaQuest collaborations. The present data are reasonably described by the predictions of the statistical parton distributions of  Basso, Bourrely, Pasechnik and Soffer\cite{Bourrely:2016} or by the chiral effective perturbation theory of Alberg and Miller,\cite{Alberg:2017ijg} also shown in Fig.~\ref{fig:dbub}.  These two calculations emphasize rather different non-perturbative mechanisms that lead to the differences in $\bar{d}(x)$ and $\bar{u}(x)$. The present data show that  $\bar d$ is greater than $\bar u$ for the entire $x$ range measured by this experiment. This provides important support for these and other non-perturbative mechanisms of the QCD structure of the proton that were disfavored by the NuSea results.

The next major step to help distinguish between the various models is to measure how much the spin and angular momentum of the antiquarks contribute to the total spin of the proton. It has long been realized that these models make rather different  predictions for the contribution of the total spin of the proton carried by the antiquarks.\cite{Chang:2014jba,Geesaman:2019} For example, meson-nucleon models predict little spin is carried by the antiquarks, the statistical model predicts the difference in spin $\Delta \bar d(x) - \Delta \bar u(x) = -\left[\bar d(x) - \bar u(x)\right]$ and chiral soliton models\cite{Pobylitsa:1999} predict $\Delta \bar d(x) - \Delta \bar u(x) = -5/3\left[\bar d(x) - \bar u(x)\right]$. Experiments are planned or underway at Fermilab, the Thomas Jefferson National Accelerator Facility, Brookhaven National Laboratory, and the European Organization for Nuclear Research to pursue this goal.\cite{Chang:2014jba,Geesaman:2019} 

These results impact the reach of a $p$-$p$ collider like the Large Hadron Collider for new physics. For example production of high mass $Z^\prime$ and $W^\prime$ particles has been shown to be dominated by light quark fusion.\cite{PhysRevD.76.115015}  Calculations with the contrasting statistical distributions (CTEQ6 distributions), which mimic the present (NuSea)  data, show that the ratio $\bar d/\bar u$ is relatively insensitive to scale in each case.  At mass scales of 4-5 TeV/c$^2$ just above current limits on $Z^\prime$  production,\cite{CMSWboson} cross sections driven by $u_v(x_1)$ $\bar d(x_2)$ fusion with $x_1 \approx x_2$ in the region of 0.3-0.4 will be enhanced based on the  present results and cross sections driven by $u_v(x_1)$ $\bar u(x_2)$ will be diminished, compared to those calculated with the central values of previous parton distributions.


\noindent{\bf Acknowledgements} We thank G. T. Garvey for important contributions to the early stages of this experiment. We also thank the Fermilab Accelerator Division and Particle Physics Division for their strong support of this experiment. 

This work was performed by the SeaQuest Collaboration whose work was supported in part by U. S. Department of Energy grants DE-AC02-06CH11357, DE-FG02-07ER41528, DE-SC0006963; U. S. National Science Foundation under grants PHY 0969239, PHY 1306126, PHY 1452636, PHY 1505458, PHY 1614456; the DP\&A and ORED at Mississippi State University; the JSPS (Japan) KAKENHI Grant Numbers 21244028, 25247037, 25800133; Tokyo Tech Global COE Program, Japan; Yamada Science Foundation of Japan; and the Ministry of Science and Technology (MOST), Taiwan. Fermilab is operated by Fermi Research Alliance, LLC under Contract No. DE-AC02-07CH11359 with the United States Department of Energy.

\noindent{\bf Author Contributions} P.E.R and D.F.G are the co-spokespersons of the experiment.  The entire SeaQuest collaboration constructed the experiment and participated in the data collection and analysis. Significant contributions to the cross section ratio analysis were made by graduate students J.D., B.K., R.E.M., S.M., D.H.M., K.Nagai, S.P., F.S., M.B.C.S., and A.S.T.  The ``extrapolation to zero intensity'' greatly benefited from the work of A.S.T. All authors reviewed the manuscript. 

\noindent{\bf Competing Interests} The authors declare no competing interests.

\noindent{\bf Data Availability} Raw data were generated at the Fermi National Accelerator Laboratory. Derived data supporting the findings of this study are available from the corresponding author upon request.

\noindent{\bf Additional Information}\\
{\bf Correspondence} should be addressed to P.E.R. (reimer@anl.gov).

\end{multicols}
\clearpage
\begin{multicols}{2}
\section*{Methods \\*[12pt]}

For the measurement of the \(pp\) and \(pd\) cross section ratios, 50.8 cm long liquid hydrogen (\num{0.069} interaction lengths) and liquid deuterium targets (\num{0.116} interaction lengths) and an empty target flask were used. The targets were interchanged every few minutes to substantially reduce time-dependent systematic effects. 

The SeaQuest spectrometer was constructed for the measurement of muon tracks in the forward region (laboratory angles less than 0.1 radian). It was composed of two magnets, four detector stations, each consisting of fast trigger detectors and drift chambers, distributed over 25 m along the beam direction with a 1 m thick iron muon identification wall before the final detector station.  The first magnet provided a 3.07 GeV/c transverse momentum kick between the target and the first detector station to enhance the acceptance for muon pairs resulting from the decay of high-mass virtual photons and to reduce the acceptance for the large background of low mass (less than 4 GeV/c$^2$) virtual photon events. It was filled with solid iron to absorb the proton beam and all other hadrons and electrons produced in either the target or this beam dump. A second magnet (with a 0.41 GeV/c transverse momentum kick) located between the first and second detector station provided for charge and momentum measurements of the muons. The iron hadron absorber between the third and fourth stations was used to establish the identification of muons.   

Opposite sign muon pairs were combined into di-muon candidates. The muons of each candidate were tracked back through the spectrometer to find if they emerged from a common vertex along the beam path and near the target. The resolution of the vertex location was about 30 cm along the beam direction compared with the 170 cm separation between the target center and the average interaction point of protons in the solid iron magnet.  Events identified as coming from the target were re-fit using the target-center vertex location, and the di-muon mass, $M$, longitudinal momentum, $P_L$, in the laboratory frame, and transverse momentum, $P_T$, were determined. With this information, the fractional momentum of the beam and target quarks participating in the reaction were calculated as
\begin{eqnarray}
x_b & = &  \frac{p_\text{target} \cdot p_\text{sum}}  {p_\text{target} \cdot \left(p_\text{beam}+p_\text{target}\right)}, \\
x_t & = &   \frac{p_\text{beam} \cdot p_\text{sum}} {p_\text{beam} \cdot \left(p_\text{beam} +p_\text{target} \right) } ,
\end{eqnarray}
where $p_\textrm{target}$ and $p_\textrm{beam}$ are the four momenta of the target and beam, respectively and $p_\textrm{sum}$ is the sum of the four momenta of the positive and negative muons. The prominent J/$\psi$ peak (resolution $\sigma$ = 0.21 GeV/c$^2$) and the requirement that events come from the target or beam dump served to calibrate the field integral of the solid iron magnet. The mass spectra are shown in Extended Data Fig.~\ref{extfig:mass}. Detailed field maps coupled with Hall probe measurements served to calibrate the second magnet. 

Various kinematic constraints were placed on the accepted events, the most important of which were to require the virtual photon mass to be greater than 4.5 GeV/c$^2$ and the primary vertex to be in the target region. The yields for each target were corrected by subtracting the appropriately weighted yield of the empty target flask.  For much of the data sample, the deuterium target had a 8.4\% $\pm$ 0.4\%  per molecule $HD$ contamination and the yields were corrected accordingly. The beam normalization and the uncertainty in the rate dependence corrections were the dominant systematic errors. Other smaller contributions include the uncertainty of the deuterium target purity, uncertainties in the target density and the contribution of the tail of the J/$\psi$ and $\psi^{\prime}$ to the di-muon mass spectrum greater than 4.5 GeV/c$^2$. 

Instantaneous fluctuations in the beam intensity while the data were being collected presented the main challenge in the data analysis.  These fluctuations occurred at the accelerator frequency of \SI{53}{\MHz} and led to a luminosity-normalized rate-dependent variation in the yield of events from the deuterium and hydrogen targets from a number of different sources.  Several approaches were considered to account for this variation.  Very generally, the simplest approach would have been to reject any event produced when the accelerator was above a certain, arbitrary (fairly low) threshold and absorb the remainder of the effect into a systematic uncertainty.  This would have had a significant impact on the statistical significance of the data.  A second approach would have been to model each individual effect in Monte Carlo, then parameterize individual effects, and finally apply the combined  parameterizations to the data.  The systematic uncertainty would need to account for the accuracy of the model and for any still unknown effects.  For the present data, a third method was chosen that allowed the full statistical power of the data to be kept without requiring that each and every effect of the intensity variation to be modeled.

This method considers only the final result--the ratio of event yields between the two targets--as a function of intensity. For each bin in $x_t$, the cross section ratio was plotted as a function of the instantaneous beam intensity when that event occurred, as illustrated in Extended Data Fig.~\ref{extfig:fits}.  The effect of the intensity dependence on the final result could then be parameterized from the measured data and then extrapolated to zero intensity.  The simplicity of this method is that the data alone are being used to measure and correct for the intensity dependence.  Since the bin boundaries in $x_t$ are arbitrary relative to the beam intensity, a smooth, common parameterization for the intensity dependence is to be expected.  A variety of parametric forms were compared to the data.  One such form is
\begin{equation}
\frac{Y_D\left(x_{t},I\right)} {2 Y_H\left(x_{t},I\right)} = R_{x_{t}} + a I + b  I^2,
\label{eq:fitform}
\end{equation}
where  is the $Y_{D\left(H\right)}$ luminosity-normalized, empty target subtracted yield of events from the deuterium (hydrogen) target.  In this form, $a$ and $b$ are parameters of the fit that are common to all $x_t$ bins describing the intensity, $I$, dependence, and $R_{x_{t}}$ is the zero intensity intercept for that bin.  The intercepts resulting from the simultaneous fit of all $x_t$ bins gives the cross section ratio $\sigRatio$ for each bin.  The common intensity parameters, $a$ and $b$, correlate $\sigRatio$ for all bins and are also determined in the simultaneous fit.  Other forms were also studied, including, for example
\begin{equation}
\frac{Y_D\left(x_{t},I\right)} {2 Y_H\left(x_{t},I\right)} = R_{x_{t}} + \left(a_0 + a_1x_t\right) I + \left(b_0+b_1x_t\right)  I^2,
\label{eq:fitform2}
\end{equation}
which allows for an $x_t$ correlated intensity dependence. An example of a less conventional extrapolation form that was considered is
\begin{equation}
\frac{Y_D\left(x_{t},I\right)} {2 Y_H\left(x_{t},I\right)} = R_{x_{t}} \cos{\left(\frac{I}{a_0+a_1x_t}\right)}.
\end{equation}
In addition, constraining either $a$ or $b$ values to zero and thus eliminating the $I$ or $I^2$ dependence was explored.  Using the Akaike information criterion, to avoid over parameterization, the form given in Eq.~\ref{eq:fitform} was chosen for the extrapolation.  The resulting fits from three representative $x_t$ bins are shown in Extended Data Fig.~\ref{extfig:fits}. A comparison with a fit to the parameterization in Eq.~\ref{eq:fitform2} was used to estimate the systematic uncertainties. The covariance matrix for the intercepts, $R_{x_{t}}$ resulting from the fit to Eq.~\ref{eq:fitform} is 
\begin{equation}
\left(
\begin{array}{ c   c  c  c  c  c  }
 2.70 & 1.19 & 1.15 & 1.20 & 1.09 & 1.16 \\
 1.19 & 1.87 & 1.25 & 1.31 & 1.19 & 1.26 \\
 1.15 & 1.25 & 1.79 & 1.25 & 1.15 & 1.21 \\
 1.20 & 1.31 & 1.25 & 2.14 & 1.20 & 1.27 \\
 1.09 & 1.19 & 1.15 & 1.20 & 2.49 & 1.16 \\
 1.16 & 1.26 & 1.21 & 1.27 & 1.16 & 4.06 \\
\end{array}
\right) \times 10^{-3}.
\label{eq:correlation}
\end{equation}
The same technique was independently applied to the data binned in transverse momenta, $P_T$, and mass, $M$.

The cross section ratios shown in Fig.~\ref{fig:crRatio} and listed in Tab.~\ref{tab:crRatio} are not corrected for acceptance.  To compare any calculation with the present data, it is necessary to consider the SeaQuest spectrometer's acceptance in $x_b$.  The appropriate theoretical cross section ratio may be calculated for a $x_t$ bin $i$ as 
\begin{equation}
\left(\frac{\sigma_D}{2\sigma_H}\right)_i = \frac{\sum{A_{ij}\sigma_D^{\text{calc}}\left(x_t,x_b,M\right)}}{2 \sum{A_{ij}\sigma_H^{\text{calc}}\left(x_t,x_b,M\right)}},
\end{equation}
where the subscript $j$ denotes the $j^{\text{th}}$ $x_b$ sub-bin of the $i^{\text{th}}$ $x_t$ bin, and $A_{ij}$ is the acceptance for that bin, tabulated in Extended Data Tab.~\ref{exttab:acc}.  Finally, $\sigma_{D(H)}^{\text{calc}}\left(x_t,x_b,M\right)$ is the calculated cross section, where the dependence on $x_t$, $x_b$, and $M$ has been made explicit.  SeaQuest used code for the NLO calculation of $\sigma_{D(H)}^{\text{calc}}\left(x,M\right)$ provided by W. K. Tung of CTEQ.
\end{multicols}

\clearpage
\onecolumn

\section*{Extended Data}

\setcounter{figure}{0}
\setcounter{table}{0}
\renewcommand{\tablename}{Extended Data Tab.}
\renewcommand{\figurename}{Extended Data Fig.}

\begin{figure*}[!b]
	\centering
		\caption{\linespread{1}\selectfont{}{\bf Comparison of NuSea and SeaQuest to NLO calculations.} \label{extfig:nloPDF} }
		\includegraphics[width=0.75\textwidth]{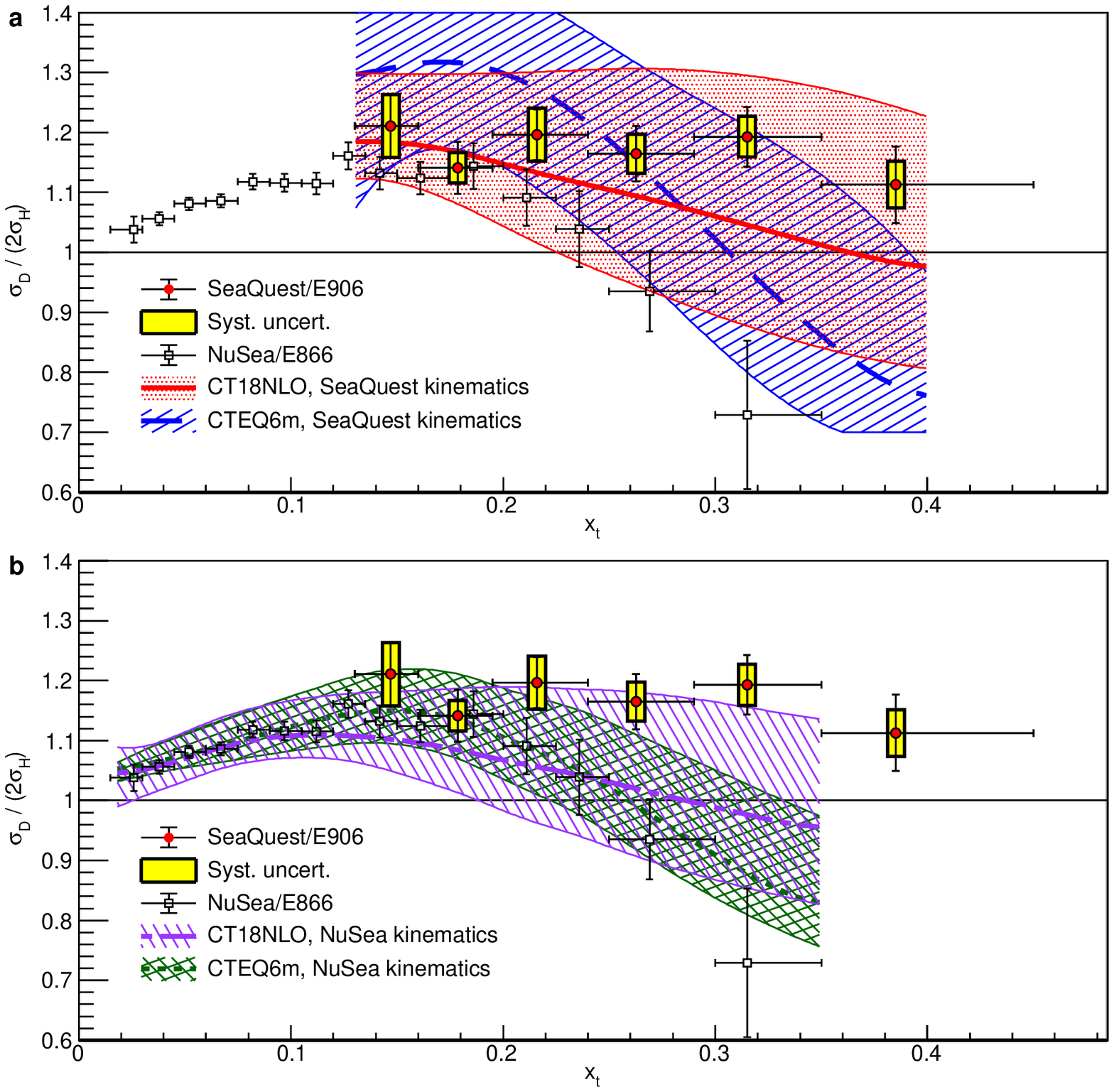}
	\flushleft
		\small{A comparison of the data from the present result and the NuSea measurements with NLO calculations made at the integrated kinematics of SeaQuest (a) and average kinematics NuSea (b) based on the CT18 and CTEQ6m parton distributions. Events in SeaQuest were produced by a \SI{120}{\GeV} proton beam while NuSea's were from an \SI{800}{\GeV} beam.  In addition, the spectrometers, while similar in concept, had different acceptances.  As a consequence, the cross section ratios, which convolve $x_t$ with $x_b$, are expected to differ because of their distinct distributions in accepted $x_b$. These kinematic effects can clearly be seen by the difference in the curves.  An acceptance table analogous to Extended Data Tab.~\ref{exttab:acc} was not available for NuSea, so those calculations used $\langle x_t \rangle$, $\langle x_b \rangle$, and $\langle M \rangle$ of the NuSea data. Both CTEQ6m and CT18 have included the NuSea data in their global analysis, so calculations based on those PDFs are expected to agree better with the NuSea data. The red (violet) curve in the upper (lower) plot is the same as in the main paper's Fig.~\ref{fig:crRatio} upper plot and repeated here for comparison.}

\vspace*{1cm} 

	\centering
		\caption{\linespread{1}\selectfont{}{\bf Extrapolation to zero intensity.} \label{extfig:fits}}
		\includegraphics[width=0.75\textwidth]{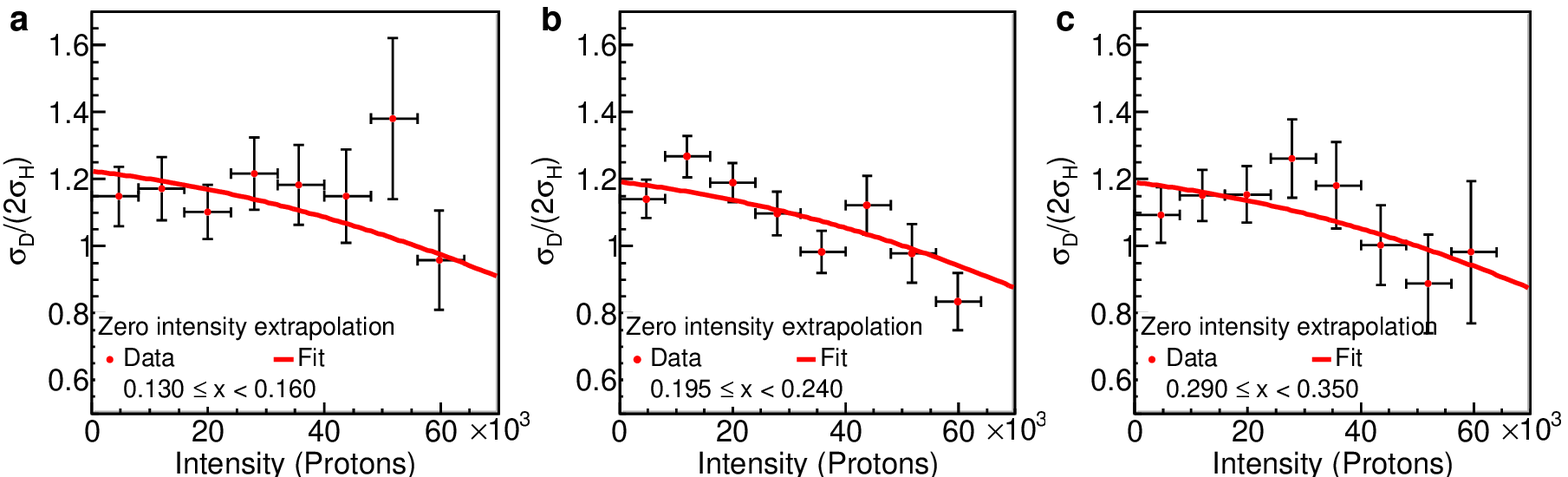}
	\flushleft
		\small{Extrapolation to zero intensity fits for representative $x_t$ bins ($0.13 \le x_t < 0.16$, $0.195 \le x_t < 0.240$, and $0.290 \le x_t < 0.350$).  The $I$ (intensity) and $I^2$ coefficients are common to all bins.  The $\chi^2/\text{dof} = 38.7/40$ for the simultaneous  fit of all $x_t$ bins. }
\end{figure*}

\begin{figure*}
	\centering
		\caption{\linespread{1}\selectfont{} {\bf Reconstructed invariant mass spectra.}  \label{extfig:mass} }
		\includegraphics[width=0.75\textwidth]{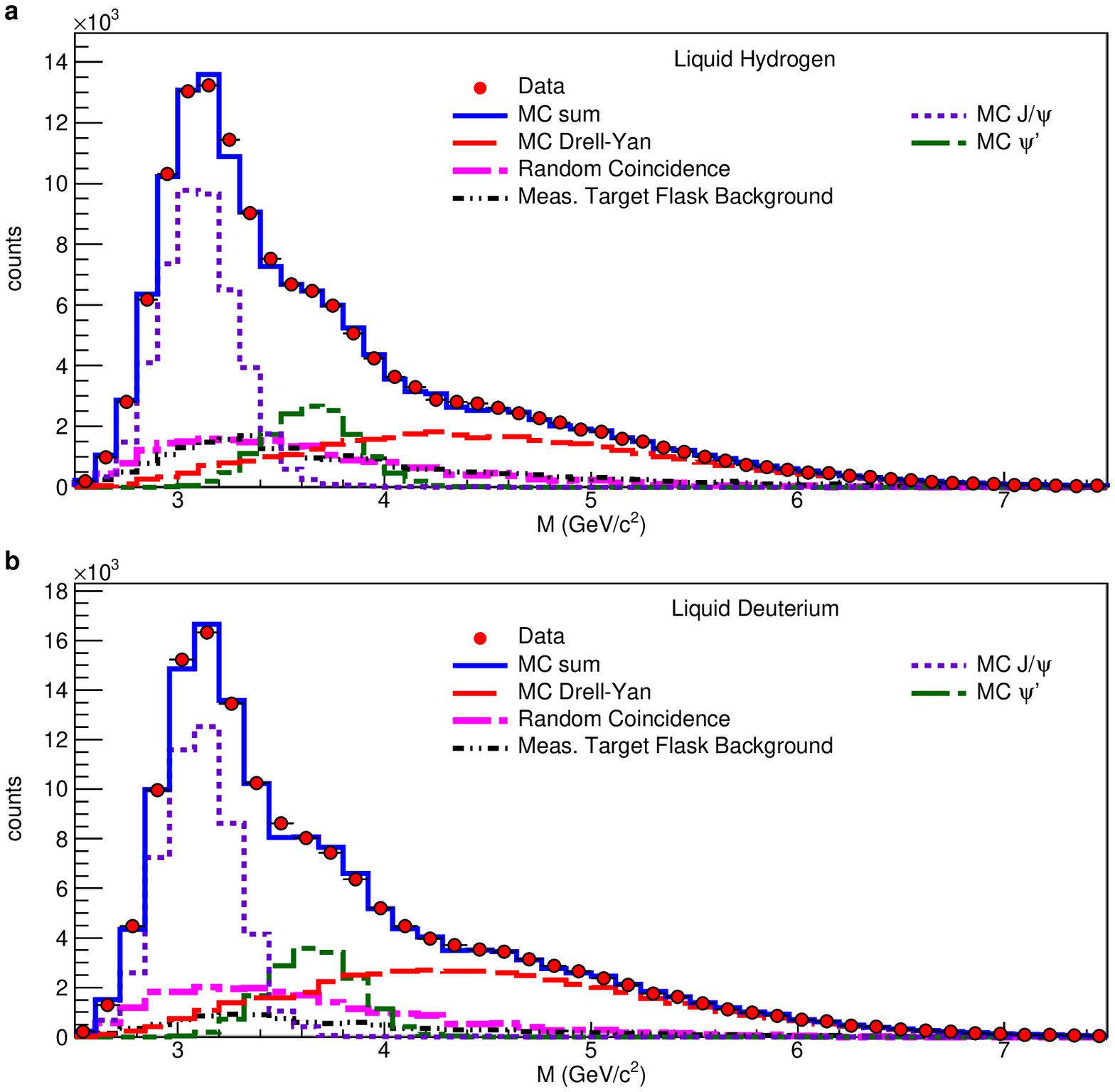}
	\flushleft
		\small{The reconstructed muon pair invariant mass spectra for the liquid hydrogen (a) and liquid deuterium (b) targets.  In the lower mass region, the predominant signal is produced by $J/\psi\rightarrow\mu^+\mu^-$decay, followed by the $\mu^+\mu^-$ decay of the $\psi^\prime$.  The prominence of the $J/\psi$ provides a calibration point for the absolute field of the solid iron magnet.  At invariant masses above 4.5 GeV/c$^2$ the Drell-Yan process becomes the dominant feature.   The data are shown as red points.  Additionally, Monte Carlo (MC) simulated distributions of Drell-Yan, $J/\psi$, and $\psi^\prime$ along with measured random coincidence and empty target backgrounds are shown.  The sum of these is shown in the blue solid curve labeled MC sum.  The normalizations of the Monte Carlo and the random background were from a fit to the data. }
\end{figure*}

\begin{table*}
    \caption{\linespread{1}\selectfont{}{\bf Ratios of $\mathbf{\sigma_D}$ to $\mathbf{2\sigma_H}$ as a function of $\mathbf{P_T}$.} \label{exttab:pt}} \vspace{6pt}
      
      \begin{tabularx}{\columnwidth}{cYr@{ $\pm$ }c@{ $\pm$ }l Y}
      \hline
      \minitab[c]{$P_T$ \\ $\left(\text{GeV/c}\right)$ } & 
      \minitab[c]{$ \langle P_T \rangle $ \\ $ \left(\text{GeV/c}\right) $ } & 
      $\displaystyle{\frac{\sigma_D}{2\sigma_H} }$ & 
        stat. &
        syst. & 
        \minitab[c]{$P_T$ \\ $\left(\text{GeV/c}\right)$ }\rule[-1em]{0pt}{2.5em} \\ \hline
      $0.0 - 0.3 $ & $ 0.198 $ & $ 1.137 $ & $ 0.049 $ & $ 0.061 $ & $ 0.161$ \pstrut \\ \hline
      $0.3 - 0.5 $ & $ 0.405 $ & $ 1.174 $ & $ 0.045 $ & $ 0.052$ & $ 0.177$ \pstrut \\ \hline
      $0.5 - 0.7 $ & $ 0.599 $ & $ 1.209 $ & $ 0.046 $ & $ 0.038$ & $ 0.188$ \pstrut \\ \hline
      $0.7 - 0.9 $ & $ 0.797 $ & $ 1.210 $ & $ 0.046 $ & $ 0.045$ & $ 0.194$ \pstrut \\ \hline
      $0.9 - 1.2 $ & $ 1.035 $ & $ 1.130 $ & $ 0.043 $ & $ 0.037$ & $ 0.198$ \pstrut \\ \hline
      $1.2 - 1.5 $ & $ 1.330 $ & $ 1.287 $ & $ 0.061 $ & $ 0.094$ & $ 0.201$ \pstrut \\ \hline
      $1.5 - 1.8 $ & $ 1.625 $ & $ 1.087 $ & $ 0.078 $ & $ 0.099$ & $ 0.206$ \pstrut \\ \hline
      $1.8 - 2.3 $ & $ 1.915 $ & $ 0.838 $ & $ 0.095 $ & $ 0.162 $ & $ 0.204$ \pstrut \\ \hline
      \end{tabularx}
      
      \vspace{6pt}
      \small{Ratios of $\sigma_D$ to $2\sigma_H$ with their statistical  and systematic uncertainties as a function of transverse momentum, $P_T$.  The cross section ratios are defined as the ratio of luminosity-corrected yields from the hydrogen and deuterium targets. The final column, $\delta P_T$ is the experimental resolution in $P_T$ as determined by Monte Carlo simulation.}
      \\
      \hrule
      \end{table*}
\begin{table*}
    \caption{\linespread{1}\selectfont{}{\bf Ratios of $\mathbf{\sigma_D}$ to $\mathbf{2\sigma_H}$ as a function of $\mathbf{M}$.} \label{exttab:mass}} 
    \vspace{6pt}
    \begin{tabularx}{\columnwidth}{cYr@{ $\pm$ }c@{ $\pm$ }l  Y}
        \hline 
        \minitab[c]{$M $                         \\ $\left(\text{GeV/c}^2\right)$} & 
        \minitab[c]{$ \langle M \rangle $ \\ $\left(\text{GeV/c}^2\right)$} & 
        $\displaystyle{\frac{\sigma_D}{2\sigma_H} }$  & 
        stat. &
        syst. & 
        \minitab[c]{$\delta M$ \\ $\left(\text{GeV/c}^2\right)$ }\\ \hline
     $  4.4 - 4.6 $ & $ 4.55 $ & $ 1.170 $ & $ 0.053 $ & $ 0.059 $ & $ 0.24$ \pstrut \\ \hline
     $  4.6 - 4.8 $ & $ 4.70 $ & $ 1.204 $ & $ 0.047 $ & $ 0.039$ & $ 0.24$ \pstrut \\ \hline
     $  4.8 - 5.0 $ & $ 4.90 $ & $ 1.202 $ & $ 0.048 $ & $ 0.039$ & $ 0.25$ \pstrut \\ \hline
     $  5.0 - 5.2 $ & $ 5.10 $ & $ 1.163 $ & $ 0.050 $ & $ 0.039$ & $ 0.26 $ \pstrut \\ \hline
     $  5.2 - 5.5 $ & $ 5.34 $ & $ 1.123 $ & $ 0.046 $ & $ 0.037$ & $ 0.26$ \pstrut \\ \hline
     $  5.5 - 6.5 $ & $ 5.89 $ & $ 1.183 $ & $ 0.043 $ & $ 0.042$ & $ 0.28$ \pstrut \\ \hline
     $  6.5 - 8.8 $ & $ 6.91 $ & $ 1.167 $ & $ 0.068 $ & $ 0.096$ & $ 0.30$ \pstrut \\ \hline
      \end{tabularx} 
      
      \vspace{6pt}
      \small{Ratios of $\sigma_D$ to $2\sigma_H$ with their statistical  and systematic uncertainties as a function of mass, $M$.  The cross section ratios are defined as the ratio of luminosity-corrected yields from the hydrogen and deuterium targets. The final column, $\delta M$ is the experimental resolution in $M$ as determined by Monte Carlo simulation.}
	\\
	\hrule
\end{table*}

\newcommand{\exht}{\rule{0pt}{2.6ex}}
 \begin{table*}[t!]
\caption{\linespread{1}\selectfont{}{\bf Spectrometer acceptance.} \label{exttab:acc}} 
\vspace{6pt}
\begin{tabularx}{\textwidth}{Y Y Y Y Y Y Y Y Y Y Y Y}
\hline
 \multirow{2}{*}{\diagbox{$x_t$}{$x_b$} }& 0.30-- & 0.35-- & 0.40-- & 0.45-- & 0.50-- & 0.55-- & 0.60-- & 0.65-- & 0.70-- & 0.75-- \exht \\
 & 0.35   & 0.40   & 0.45   & 0.50   & 0.55    & 0.60   &   0.65 & 0.70 & 0.75 & 0.80 \\ \hline 
\multirow[b]{2}{*}{0.130--} &   &   &  &   &  & 0.0007 & 0.0064 & 0.0175 & 0.0304 & 0.0370  \exht \\  
&   &  &   &   &   & 0.589 & 0.628 & 0.675 & 0.723 & 0.772  \\  
\multirow[t]{2}{*}{0.160} &   &   &   &   &   & 0.158 & 0.153 & 0.148 & 0.144 & 0.144  \\
&    &   &   &    &   &  4.54 &  4.60 &  4.68 &  4.77 &  4.92 \\ \hline 
\multirow[b]{2}{*}{0.160--} &   &   &   & 0.0007 & 0.0071 & 0.0188 & 0.0299 & 0.0366 & 0.0432 & 0.0471 \exht \\  
&   &   &   & 0.489 & 0.528 & 0.576 & 0.624 & 0.673 & 0.722 & 0.772  \\  
\multirow[t]{2}{*}{0.195} &  &   &  & 0.191 & 0.184 & 0.178 & 0.176 & 0.176 & 0.176 & 0.175  \\
&   &   &   &  4.56 &  4.63 &  4.74 &  4.91 &  5.09 &  5.27 &  5.45 \\ \hline 
\multirow[b]{2}{*}{0.195--} &   & 0.0001 & 0.0023 & 0.0105 & 0.0205 & 0.0298 & 0.0384 & 0.0456 & 0.0510 & 0.0557 \exht \\  
&  & 0.394 & 0.433 & 0.477 & 0.524 & 0.574 & 0.623 & 0.672 & 0.722 & 0.772  \\  
\multirow[t]{2}{*}{0.240} &  & 0.235 & 0.225 & 0.217 & 0.216 & 0.215 & 0.215 & 0.214 & 0.215 & 0.214  \\
&   &  4.55 &  4.65 &  4.78 &  4.99 &  5.21 &  5.43 &  5.63 &  5.84 &  6.04  \\ \hline 
\multirow[b]{2}{*}{0.240--}&   & 0.0015 & 0.0078 & 0.0176 & 0.0270 & 0.0364 & 0.0436 & 0.0499 & 0.0550 & 0.0591 \exht \\  
&  & 0.383 & 0.427 & 0.475 & 0.524 & 0.574 & 0.623 & 0.672 & 0.722 & 0.771  \\  
\multirow[t]{2}{*}{0.290} &   & 0.267 & 0.264 & 0.263 & 0.262 & 0.262 & 0.262 & 0.262 & 0.261 & 0.262  \\
&    &  4.76 &  4.99 &  5.24 &  5.50 &  5.75 &  6.00 &  6.24 &  6.46 &  6.69  \\ \hline 
\multirow[b]{2}{*}{0.290--} & 0.0002 & 0.0035 & 0.0120 & 0.0207 & 0.0298 & 0.0379 & 0.0455 & 0.0518 & 0.0544 & 0.0568  \exht \\
& 0.341 & 0.379 & 0.426 & 0.475 & 0.524 & 0.574 & 0.623 & 0.673 & 0.722 & 0.771  \\  
\multirow[t]{2}{*}{0.350}  &0.324 & 0.319 & 0.316 & 0.316 & 0.316 & 0.315 & 0.315 & 0.315 & 0.314 & 0.314  \\
&  4.95 &  5.18 &  5.46 &  5.76 &  6.05 &  6.33 &  6.60 &  6.85 &  7.10 &  7.34  \\  \hline 
\multirow[b]{2}{*}{0.350--} & 0.0006 & 0.0052 & 0.0125 & 0.0203 & 0.0268 & 0.0336 & 0.0374 & 0.0405 & 0.0415 & 0.0413  \exht \\
& 0.339 & 0.377 & 0.425 & 0.474 & 0.524 & 0.573 & 0.623 & 0.672 & 0.722 & 0.771  \\  
\multirow[t]{2}{*}{0.450} & 0.384 & 0.390 & 0.386 & 0.386 & 0.385 & 0.384 & 0.384 & 0.384 & 0.383 & 0.382  \\
&  5.38 &  5.72 &  6.04 &  6.38 &  6.69 &  7.00 &  7.29 &  7.58 &  7.85 &  8.11  \\ \hline 
\end{tabularx}

\vspace{6pt}
\small{The acceptance relative to a $4\pi$ detector and average kinematic values for bins in $x_t$ and $x_b$.  Each cell gives the acceptance (top), $\langle x_b\rangle$, $\langle x_t\rangle$, and $\langle \text{mass}\rangle$ (bottom) for each sub-bin. }
	\\
	\hrule
\end{table*}

\end{document}